\documentclass[a4paper,10pt]{article}
\usepackage{enumerate}
\usepackage{color}
\usepackage[utf8]{inputenc} 
\usepackage[english]{babel}
\usepackage[T1]{fontenc}
\usepackage{graphicx}
\usepackage{amsfonts,amssymb,amsmath,latexsym,amsthm}
\usepackage{textcomp}
\usepackage[pdftex]{hyperref}
\usepackage{geometry}
\DeclareMathOperator{\sech}{sech}

\title{Bipolar solitary wave interactions within the Schamel equation}
\author{Ekaterina Didenkulova$^{1,2}$, Efim Pelinovsky$^{1,2}$ and  Marcelo V. Flamarion$^{3}$}
\date{}

\begin{document}
\maketitle
\begin{center}

{\footnotesize 
$^{1}$Faculty of Informatics, Mathematics and Computer Science, HSE University, Nizhny Novgorod 603155, Russia.

$^{2}$ Il'ichev Pacific Oceanological Institute Far Eastern Branch Russian Academy of Sciences, Vladivostok 690041, Russia.}
\vspace{0.3cm}

{\footnotesize $^3$Unidade Acad{\^ e}mica do Cabo de Santo Agostinho, \\
UFRPE/Rural Federal University of Pernambuco, BR 101 Sul, Cabo de Santo Agostinho-PE, Brazil,  54503-900 \\
marcelo.flamarion@ufrpe.br }




\end{center}


\begin{abstract} 
Pair soliton interactions play a significant role in the dynamics of soliton turbulence. The interaction of solitons with different polarities is particularly crucial in the context of abnormally large wave formation, often referred to as freak or rogue waves, as these interactions result in an increase in the maximum wave field. In this article, we investigate the features and properties of bipolar soliton interactions within the framework of the non-integrable Schamel equation, contrasting them with the integrable modified Korteweg-de Vries equation. We examine variations in moments and extrema of the wave fields. Additionally, we identify scenarios in which, in the bipolar solitary wave interaction, the smaller solitary wave transfers a portion of its energy to the larger one, causing an increase in the amplitude of the larger solitary wave and a decrease in the amplitude of the smaller one, returning them to their pre-interaction state. Notably, we observe that non-integrability can be considered a factor that triggers the formation of rogue waves.
\end{abstract}

\section{Introduction}
The existence of coherent structures like solitons in nonlinear wave systems leads to a
problem of soliton gas or soliton turbulence, which has been in focus of investigation for the last
decade \cite{Suret:2023, Zakharov:2009, G:2021, Shurgalina:2017, GA:2016, Carbone:2016}. Such wave fields imply the presence of a large number of solitons with random
parameters which significantly determine the dynamics of the wave field. The problem of soliton
turbulence can be expanded to the breather turbulence or soliton-breather turbulence, if apart
from solitons there is also another type of waves that conserve their energy during
propagation--oscillating wave packets called breathers \cite{Akhmediev:2016, Crespo:2016, Didenkulova:2022}. The study of soliton turbulence
arose in the context of integrable equation through the inverse scattering transform equations, most known of
which are the nonlinear Schrödinger (NLS) equation, which describes time evolution of the
envelope of a quasi-monochromatic wave train \cite{Zakharov:1972} and the Korteweg-de Vries equation \cite{Zakharov:2009, Zakharov:1971}.
The interaction of the solitons in the framework of these equations is elastic, and there is no loss
of energy after such collisions.

Two-soliton collisions and their properties were extensively studied in many articles, for
example, in \cite{Pelinovsky:2013, Slunyaev:2001, Shurgalina:2018a, Shurgalina:2018b, Shurgalina:2015, Ali:2020} for KdV-like models, such as Korteweg-de Vries, modified Korteveg-de Vries
(mKdV), and extended Korteweg-de Vries (known as Gardner) equations. In the work  of Anco et al. \cite{Anco:2011}, the
collisions of two solitons were studied within a more exotic equation of KdV-type: Sasa-Satsuma
and Hirota equations. The most significant parameter, which defines the character of the wave
interaction is the polarity of the solitons. Solitons with the same polarity repel each other and the
amplitude of the resulting impulse in the moment of interaction decreases and it is smaller than
the amplitude of the largest soliton. In the case of interaction of solitons with different polarity,
the amplitude of the resulting impulse increases due to attraction of solitons. Thus, the behavior
of soliton gas depends noticeably on the polarities of solitons. The wave field consisting of
bipolar solitons (i.e. positive and negative solitons) experiences anomalously large waves
formation \cite{Didenkulova:2019, Shurgalina:2016a}. Moreover, analytically, it was shown that in the case of very specific soliton phases (or positions), all solitons (and breathers) have the potential to coalesce into a single massive rogue wave, with its amplitude being the linear sum of the amplitudes of the interacting waves \cite{Pelinovsky:2016, Slunyaev:2018}. However, in real systems, such situations should happen rarely, and the soliton interactions is a quasi-chaotic process.

There are non-integrable modifications of KdV equations, which have soliton solutions.
However their interaction is inelastic and there is additional radiation, which occurs after the
soliton interactions. Here, apart from the other equations, the number of the Schamel-like
equations, describing ion acoustic waves due to resonant electrons, nonlinear wave dynamics of
cylindrical shells, and longitudinal waves in the walls of an annular channel can be listed \cite{Schamel:1973, Schamel:2000, Zemlyanukhin:2019, Mogilevich:2023, Schamel:2017}.
Unlike the KdV equation, the nonlinear term of the Schamel equation contains a module
of the wave elevation function. This equation allows the existence of solitons with both positive and
negative polarity, thus the problem of soliton turbulence can be relevant for the wave systems described
by this equation. The article about the study of the process of positive pair soliton interaction
within the Schamel equation was published recently \cite{Flamarion:2023}. It was shown that some features of
solitary wave interaction, like the evolution of wave moments or phase shifts which acquire solitons
after interaction, are similar to ones within integrable KdV-like models due to small dispersion.
Also, it was shown previously that the dynamics of the wave field consisting of a large number
of solitons with the same polarity is very close in the case of integrable KdV-equation and non-integrable BBM-equation \cite{Dutykh:2014}.

Here, we concentrate on bipolar soliton interaction within the Schamel equation and its
comparison with the integrable mKdV equation. The structure of the present article is as follows:
Section 2 contains the description of two-soliton collision within the Schamel equation. The
comparison of the properties of two-soliton interactions for integrable and non-integrable models
is given in Section 3. The analysis of the evolution of moments of the wave fields are presented
in Section 4. Finally, there are conclusion remarks at the end of the article.

\section{Bipolar soliton interaction in the framework of the Schamel equation}
In our research, we investigate solitary wave interactions by focusing on the Schamel equation in its canonical form
\begin{equation}\label{Schamel1}
u_{t} +\sqrt{|u|}u_{x}+u_{xxx}=0.
\end{equation}
Within this equation, the variable $u$ represents the wave field at a specific position $x$ and time $t$.
\begin{figure}[h!]
\centering	
\includegraphics[scale=1]{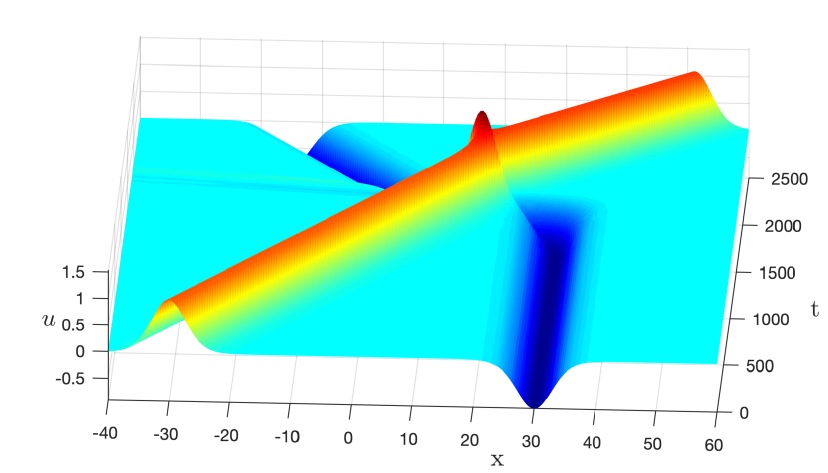}
\includegraphics[scale=1]{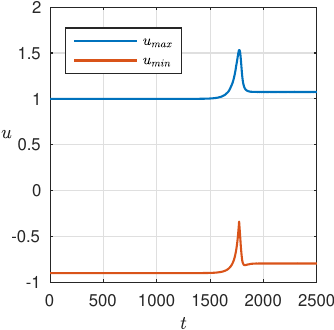}
\includegraphics[scale=1]{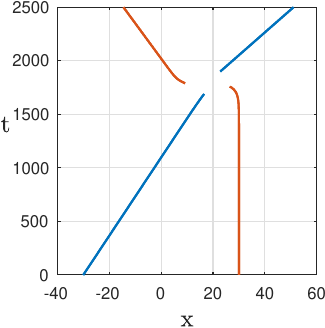}
\caption{ Top: Overtaking collision of two solitary waves for the Schamel equation. Bottom: (Left) The maximum value of the solution as a function time. Right: The crest trajectories of the solitary waves. Parameters $A_1$ = 1.0 and $A_2 = -0.9$.} 
\label{Fig1}
\end{figure}

The Schamel equation (\ref{Schamel1}) supports solitary waves as its solutions. These solitary waves can be described by the following expressions
\begin{equation}\label{solitary}
u(x,t) = a\sech^{4}\left(k(x-ct)\right), \mbox{ where }  c = \frac{8\sqrt{|a|}}{15} \mbox{ and } k = \sqrt{\frac{c}{16}}.
\end{equation}
Here, $a$ stands for the amplitude of the solitary wave, which can be positive or negative. The parameter $c$ denotes the speed of the solitary wave and $k$ characterizes its wavenumber.

We set initial wave field as linear superposition of two solitons, taking into account that
solitons are located far enough from each other soliton field at $t=0$ 
\begin{equation}
u(x,0)=u_{1}(x)+u_{2}(x),
\end{equation}
where $u_{1}$, $u_{2}$ represent one-soliton solutions with amplitudes $a_{1}$ and $a_{2}$ respectively. Solutions are computed numerically using the standard pseudoespectral method with integrating factor presented in \cite{Trefethen:2001}. The time advanced computed through the Runge-Kutta forth-order method.
\begin{figure}[h!]
\centering	
\includegraphics[scale=1]{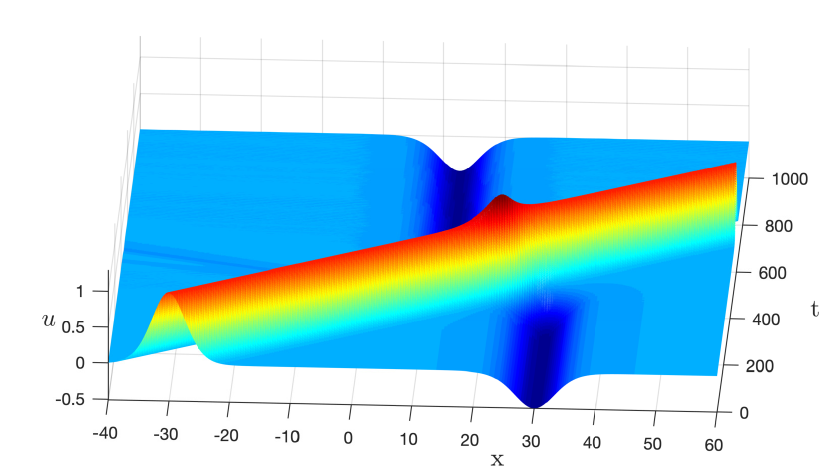}
\includegraphics[scale=1]{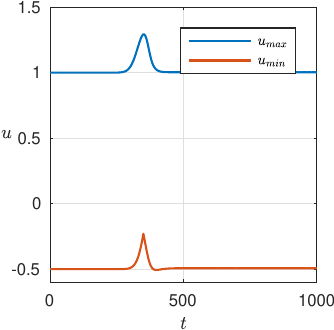}
\includegraphics[scale=1]{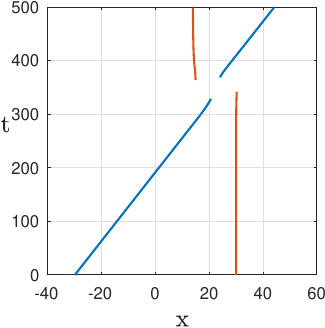}
\caption{ Top: Overtaking collision of two solitary waves for the Schamel equation. Bottom: (Left) The maximum value of the solution as a function time. Right: The crest trajectories of the solitary waves. Parameters $A_1$ = 1.0 and $A_2 = -0.5$.} 
\label{Fig2}
\end{figure}

The properties of pair soliton interactions with and can be found in Figure \ref{Fig1}. The bottom
of this figure (left plot) demonstrates the evolution of the maximum of the wave field. It can be
clearly seen that before interaction, the value of is equal to the value of the amplitude of the
largest soliton; at the moment of interaction, it increases up to the value 1.53, however after
the separation, does not return to its original value and remains at the level of 1.09. It means that a positive soliton increases its energy after the soliton collision. Consequently, the negative
one loses its energy as it is clearly seen in  Figure \ref{Fig1} (bottom left), where the value of increases
after the soliton interaction and becomes equal to -0.8. There is no radiation which attenuates
both solitons. In comparison with the integrable mKdV case, it is inelastic collision with energy
transfer from smaller to larger soliton. Thus, it can play a dramatic role in the soliton gas, when
the energy concentrates in the biggest soliton which grows with time. It can be considered as a
new mechanism of rogue wave generation due to non-integrability. In fact, when the solitons
have the same polarity \cite{Flamarion:2023}, the dynamics of the wave field resembles the KdV-collisions except
for the small radiation produced during the collision. In other words, the amplitudes are almost
conserved.

Another intriguing finding is that the angle formed with the positive axis Ox and the
soliton trajectory changes after the collisions (Figure \ref{Fig1}, bottom right). For instance, the change for
the negative soliton is $90^o ?103^o = ?13^o$ , while for the positive soliton is $88^o-87^o=1^o$. There is a big difference in comparison
with the mKdV equation where only a phase-shift occurs after the collision and the trajectories have the same angle.

The features of soliton interactions depend a lot on the amplitudes of the solitons.
Another example of pair soliton interactions with and is presented in Figure \ref{Fig2}. Here, the value of
almost returns to its original value, and the angle between the positive axis Ox and the soliton
trajectory does not change after the collisions. The value of also restores after the solitons
become separated.
\begin{figure}[h!]
\centering	
\includegraphics[scale=1.2]{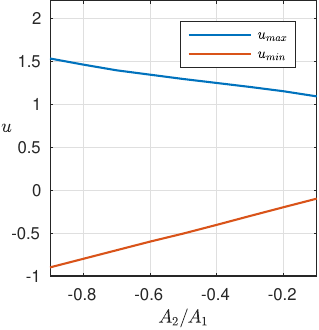}
\includegraphics[scale=1.2]{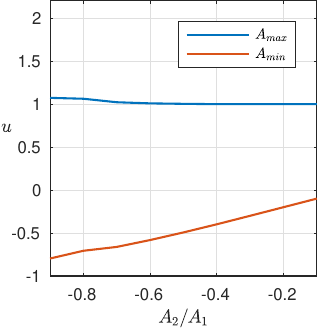}
\caption{Left: Extrema of wave fields consisting of bipolar solitons described by the Schamel equation as a function of the ratio $A_{2}/A_{1}$. Here, $u_{max}$ denotes the maximum vales of the wave field and  $u_{min}$ its minimum value over the solitary wave interaction. Right: The amplitude of the solitary waves after the interaction as a function of the ratio $A_{2}/A_{1}$.}
\label{Fig3}
\end{figure}

To study the dependence of extrema (maximum and minimum) of wave fields against the
ratio of the amplitudes of the interacted solitons, as well as the values of amplitudes of the
largest and smallest solitons after the soliton interaction, Figure \ref{Fig3} is analyzed. The amplitude
of the largest soliton before the soliton collision was kept constant and equal to 1. The amplitude
of the second soliton was changed from -0.9 to -0.1. It is observed that the minimum of the wave
field for all considered ratios $A_2/A_1$ is equal to the amplitude of the smallest soliton and does not
fall below this value over time. Thus, the straight red line can be seen in the left plot of  Figure \ref{Fig3}.
The value of $u_{min}$ before and after the soliton interaction is not the same for all ratios $A_2/A_1$. This
difference can be seen for amplitudes $A_2$ less than -0.7 (see red line on the right plot of  Figure \ref{Fig3}). It
means that the negative soliton transfers part of its energy to the positive one only when their
amplitudes are comparable in modulus. Accordingly, this effect does not appear when the
amplitude of the negative soliton is considerably smaller than the amplitude of the large positive
soliton. A similar conclusion can be made for the wave field maxima: the amplitude of the
largest soliton after the soliton interaction exceeds its initial value (which is equal to 1) only for
ratio $|A_2/A_1|$ bigger than 0.7.


\section{Comparison of the process of the soliton interaction within the Schamel and the mKdV
equation}
In the present Section, the qualitative differences between the interaction of bipolar solitons
within the non-integrable Schamel equation and integrable modified Korteweg-de Vries equation
are analyzed.

The mKdV equation 
\begin{equation}\label{mKdV}
u_{t} +u^{2}u_{x}+u_{xxx}=0.
\end{equation}
also admits solitary wave as solutions described by the formulas
\begin{equation}\label{solitaryKdV}
u(x,t) = a\sech(\left(k(x-ct)\right), \mbox{ where }  v = \frac{a^2}{6} \mbox{ and } k =\sqrt{\frac{a^2}{6}}.
\end{equation}

\begin{figure}[h!]\
	\centering	
	\includegraphics[scale =1]{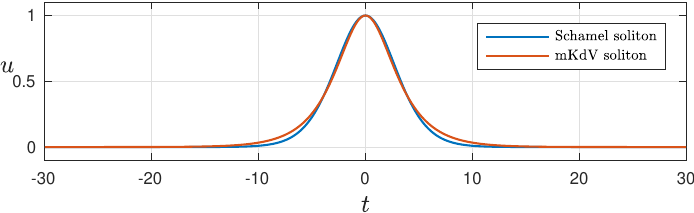}
	\includegraphics[scale =1]{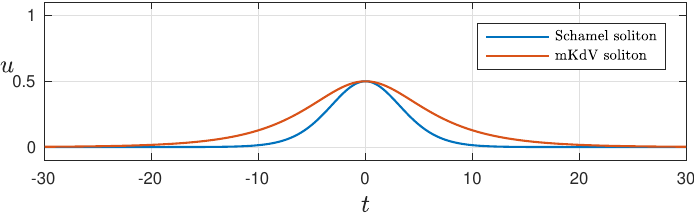}
	\caption{ Soliton profile of the Schamel equation (\ref{solitary})  and the KdV equation (\ref{solitaryKdV}).}
	\label{Solitons}
\end{figure}

\begin{figure}[h!]
\centering	
\includegraphics[scale=1]{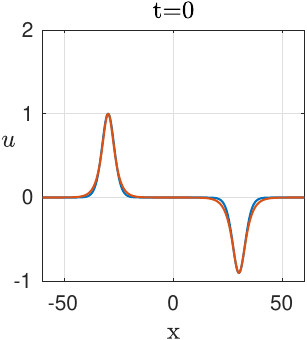}
\includegraphics[scale=1]{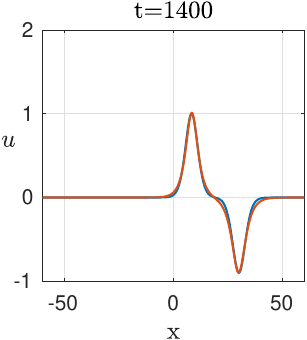}
\includegraphics[scale=1]{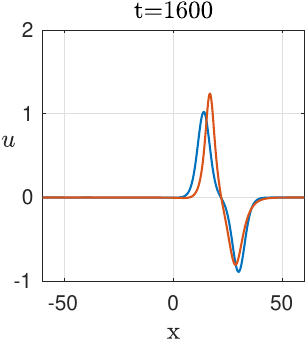}
\includegraphics[scale=1]{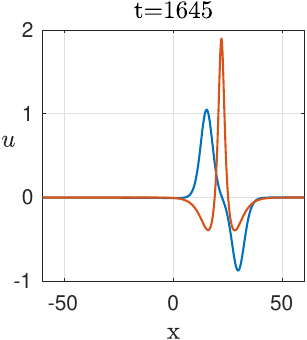}
\includegraphics[scale=1]{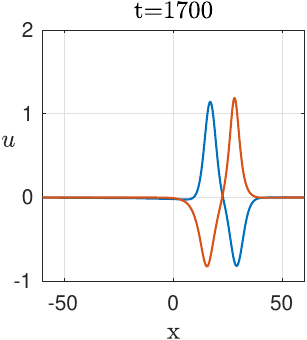}
\includegraphics[scale=1]{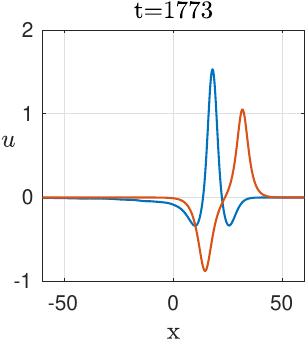}
\includegraphics[scale=1]{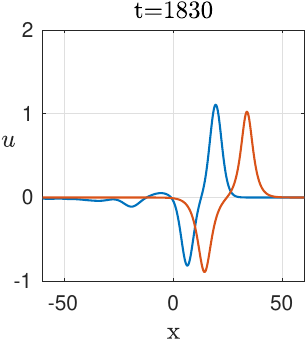}
\includegraphics[scale=1]{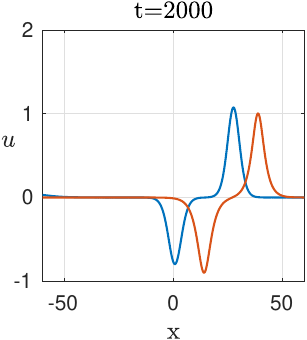}
\includegraphics[scale=1]{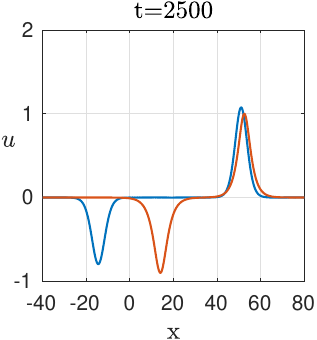}
\caption{A series of snapshots of the interaction of the solitary waves for the Schamel (in blue) and mKdV equation (in red). Parameters $A_1$ = 1.0 and $A_2 = -0.9$.} 
\label{Fig5}
\end{figure}

\begin{figure}[h!]
\centering	
\includegraphics[scale=1]{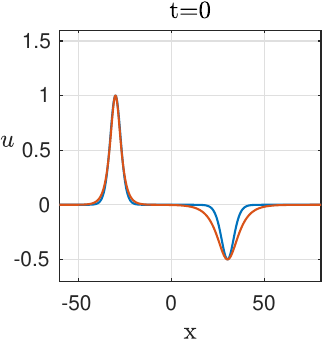}
\includegraphics[scale=1]{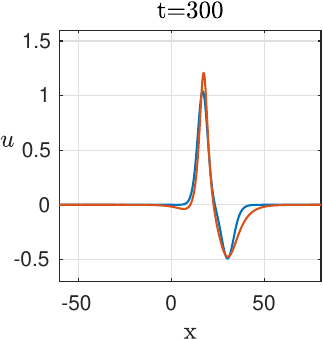}
\includegraphics[scale=1]{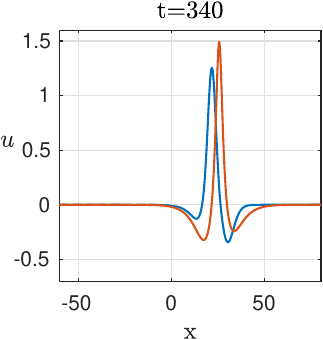}
\includegraphics[scale=1]{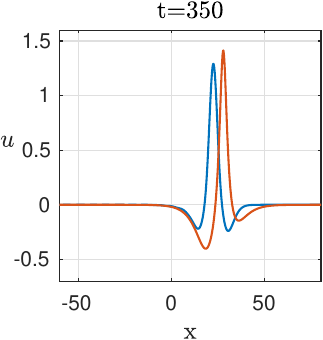}
\includegraphics[scale=1]{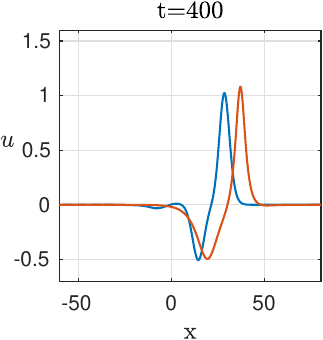}
\includegraphics[scale=1]{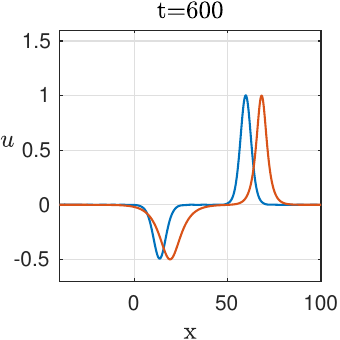}
\caption{A series of snapshots of the interaction of the solitary waves for the Schamel (in blue) and mKdV equation (in red). Parameters $A_1$ = 1.0 and $A_2 = -0.5$.} 
\label{Fig6}
\end{figure}

The soliton solutions of the Schamel equation are much narrower than the mKdV solitons with
small amplitudes (Figure \ref{Solitons}). Large solitons within the Schamel equation are a bit wider at the top
of the soliton and narrower at the bottom in comparison with the mKdV solitons. Schamel
solitons propagate faster than mKdV ones. These differences in shapes and velocities contribute
to the difference in the interaction processes of solitons.
The process of two soliton interactions in integrable mKdV model and non-integrable
Schamel equation is presented in Figures \ref{Fig5} and \ref{Fig6} for the simulations presented in Figure \ref{Fig1}, and Figure \ref{Fig2} correspondingly. The interaction of
bipolar solitons leads to the increase in the amplitude of the resulting impulse. However, within
the mKdV model, the resulting amplitude presents the linear superposition of the modules of
amplitudes of the initial solitons. The amplitude of the resulting impulse within the Schamel
equation is less due to the additional radiation which occurs because of the non-integrability of
this equation and inelastic interaction of solitons. Also, in the mKdV model, the interacting
solitons restored their shapes and amplitudes after the interaction, and the biggest wave which
can form in the soliton gas has the amplitude equal to the sum of all solitons; this scenario
requires a very specific soliton phases and order of solitons \cite{Pelinovsky:2016, Slunyaev:2018}. In the framework of the
Schamel equation, when solitons have close amplitudes by modulus (Figure \ref{Fig5}), the scenario, on
the one hand, is less extreme than the mKdV case because the resulting impulse is less than in
the mKdV case. On the other hand, the amplitude of the largest soliton increases after the
interaction. It may lead to the consumption of the energy of the largest soliton in soliton gas and,
as a result, it may be considered as a new mechanism of a freak or rogue wave formation due to
the non-integrable effects. When the initial amplitudes of the solitons significantly differ from
each other, the resulting impulse in the Schamel equation is again less than in the mKdV case,
but not significantly, because the non-integrable radiation is small in this case (Figure \ref{Fig6}). Moreover,
after the interaction, the solitons have similar shapes as they had before the interaction.

Figures \ref{Fig5} and \ref{Fig6} are plotted in the Schamel time, and the corresponding mKdV time is computed as
\begin{equation}
t_{m}=\frac{c_1-c_2}{v_1-v_2}.
\end{equation}
Here, $c_1$ represents the positive Schamel soliton speed, while $c_2$ corresponds to the negative Schamel soliton speed. Similarly, $v_1$ denotes the speed of positive mKdV solitons, and $v_2$ represents the speed of negative mKdV solitons.

\section{Moments of the wave fields}
\begin{figure}[h!]
\centering	
\includegraphics[scale=1.2]{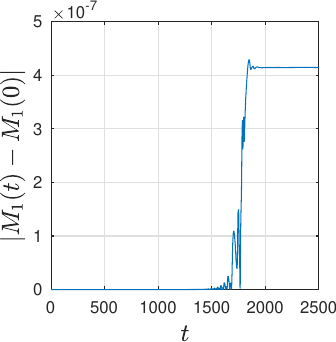}
\includegraphics[scale=1.2]{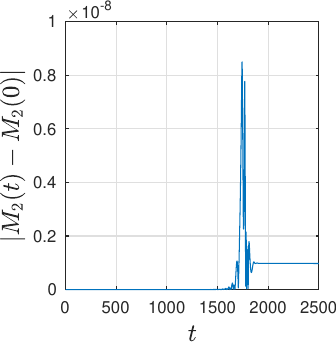}
\caption{ The variation of moments $M_1$ and $M_2$ as functions of time during interaction of the solitary waves of the Schamel equation for the collisions displayed in Figure \ref{Fig1}.} 
\label{Fig4}
\end{figure}

With aim to understand the contribution and the influence of pair interactions on the statistical
moments of the wave field, the following integrals which correspond to four statistical moments
were studied
\begin{equation}
M_{n}(t)=\int_{-\infty}^{+\infty}u^{n}(x,t), \mbox{ for $n=1,2,3,4.$}
\end{equation}
The first invariant $(n=1)$ is known as the Casimir invariant or the mass invariant. It
characterizes the mass or the total ``amount" of the wave field at any given time $t$. The second
invariant $(n=2)$ is the momentum invariant. These invariants play a crucial role in
evaluating the accuracy and reliability of numerical methods employed to solve the Schamel
equation (\ref{Schamel1}). Their evolution within the Schamel equation is presented in Figure \ref{Fig4}. It can be seen that the accuracy of the calculations is quite high, and the first and second integrals are preserved with the
accuracy equal to $10^{-7}$ and $10^{-8}$.

\begin{figure}[h!]
\centering	
\includegraphics[scale=1.2]{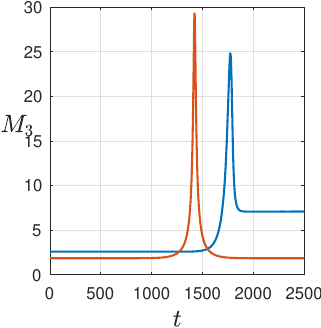}
\includegraphics[scale=1.2]{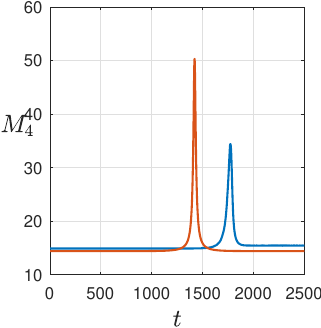}
\includegraphics[scale=1.2]{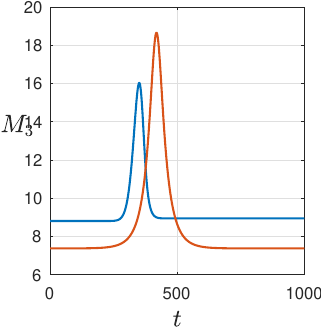}
\includegraphics[scale=1.2]{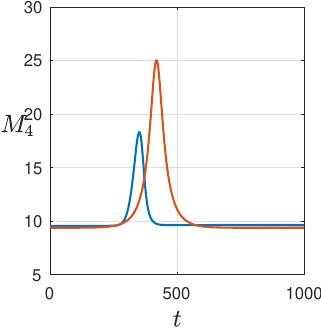}
\caption{ Moments $M_3$ and $M_4$ as functions of time during interaction of the solitary waves of the Schamel equation (in blue) and mKdV equation (in red) for the collisions displayed in Figures \ref{Fig1} (top) and \ref{Fig2} (bottom) respectively.} 
\label{Fig8}
\end{figure}

The third and fourth moments corresponding to the skewness and the kurtosis in the
theory of turbulence vary in time and change their values during the soliton interaction. Their
evolution and comparison for the two equations are shown in Figure \ref{Fig8} (top plot - in case of the simulation displayed in Figure \ref{Fig1})
bottom plot - in case of the simulation displayed in Figure \ref{Fig2}). These moments increase in the moment of soliton interactions due to
the increase in amplitude of the resulting impulse. The peaks of the moments are predictably
higher within the mKdV equation than in the Schamel equation, because the growth of impulses
is larger in the mKdV equation as it was shown in the previous section. By analogy with the
maxima of wave fields, in the integrable mKdV equation, due to elastic soliton interactions, the
values of the third and fourth moments before and after the interaction are identical. The same
conclusion can be made for the Schamel equation in case of large difference in amplitudes of interacted solitons. The radiation due to the non-integrability of this equation is small enough.
However, when the amplitudes of solitons are close by modulus ($A_2/A_1$ is bigger than 0.7), the
value of the moments after the interaction becomes bigger than before the interaction. It means
that in case of the mKdV soliton interaction, the amplitudes of interacted solitons are higher than
in the Schamel case. However, if in the Schamel equation, a large time of evolution of the wave
field consisting of a big number of solitons is considered, the accumulation of energy in large
solitons can contribute to the emergence of unexpectedly large waves in the system. Thus, non-integrability can be considered as a factor provoking the rogue wave formation.

\section{Conclusion}
The present article is devoted to the study of the process of collision of two solitons with
different polarities, which is considered as an elementary act of soliton turbulence. The study is
carried out numerically within the non-integrable Schamel equation. This equation has a soliton
solution similar to the known integrable KdV-like equations. However, due to the non-integrability, the solitons interact inelastically and radiation occurs. It was shown that in the
process of bipolar soliton interaction, the amplitude of the resulting impulse increases, but it is
less than the sum of soliton amplitudes before the interaction. This fact differs these interactions
from soliton interactions within the integrable models, where the amplitude of the resulting
impulse is equal to the linear superposition of soliton amplitudes before interaction. However,
the most important finding of the present work is the fact that after the bipolar soliton interaction
within the Schamel equation, the smaller soliton may transfer a part of its energy to the larger
one, so the amplitude of the larger soliton becomes higher and the amplitude of the smaller one
becomes lower as it was before the interaction. This effect is especially evident in the case of
interaction of bipolar solitons with amplitudes close in magnitude. Thus, large solitons may
accumulate energy. This is particularly important in the contest of soliton turbulence or soliton
gas, when the system consists of a large number of solitons, which interact with each other
repeatedly.

Previously, it was shown that in the integrable KdV-like models, the bipolar soliton
interactions can be a mechanism of rogue wave formation since the optimal focusing promotes
the formation of the wave with an amplitude equal to the sum of amplitudes of interacted
solitons even if their number is bigger than two. Within the Schamel equation, this effect is less
pronounced due to the non-integrable radiation which occurs after the interaction. However,
the energy accumulation in the largest solitons may contribute to the rogue wave formation in the
case of long-term wave dynamics.

\section{Acknowledgements}
E.P. and E.D. are supported by Laboratory of Nonlinear Hydrophysics and Natural Disasters of the V.I. Il'ichev Pacific Oceanological Institute, grant from the Ministry of Science and Higher Education of the Russian Federation, agreement number 075-15-2022-1127 from 01.07.2022.

	\section*{Declarations}
	
	\subsection*{Conflict of interest}
	The authors state that there is no conflict of interest. 
	\subsection*{Data availability}
	
	Data sharing is not applicable to this article as all parameters used in the numerical experiments are informed in this paper.

\end{document}